\documentstyle[aps,preprint]{revtex}
\begin{document}

\title { Finite coherent length and multi-pion correlation effects on 
two-pion interferometry}
\author{Q. H. Zhang$^a$, X. Q. Li$^{b,c}$, C. S. Gao$^{b,d,f}$ and W. Q. Chao$^{b,e,f}$}
\address{a Institut f\"ur Theoretische Physik, Universit\"at Regensburg,
D-93040 Regensburg, Germany\\
b  China Center of Advanced Science and Technology(World Laboratory),
P.O. Box 8730,Beijing 100080, P.R. China\\
c Physics Department, Nankai University, Tianjing 300071, China\\
d Physics Department, Peking University, Beijing 100871, China\\
e Institute of High Energy Physics, Academia Sinica, P.O. Box 918(4), 
Beijing 100039, China\\
f Institute of Theoretical Physics, Academia Sinica, P.O. Box 2735, 
Beijing 100080, China}
\vfill
\maketitle

\begin{abstract}
The effects of multi-pion correlations and finite coherent length 
on two-pion interferometry are studied. It was shown that 
as the pion multiplicity and coherent length become larger,
the apparent radius and the apparent coherent parameters derived from two-pion 
interferometry become 
smaller.  The influence of the 
coherent length on the effective temperature is 
discussed.
 
\end{abstract}

PACS number(s): 25.75 Dw, 11.38 Mh, 11.30 Rd

Two-pion Bose-Einstein(BE) correlation is widely used in high energy 
heavy-ion collisions to provide the information of the space-time
structure, degree of coherence and dynamics of the region where 
the pions were produced\cite{HBT,GGLP,BGJ,GKW,ZL,Heinz,APW}.   
Ultrarelativistic hadronic and nuclear collisions provide 
the environment for creating dozens, and in some cases hundreds, 
of pions\cite{NA35,E802,NA44}.   The bosonic nature of the pion should 
affect the single and two-pion spectra and distort the two-pion 
correlation 
function\cite{WC84,Zajc87,Pratt93,PGG90,Cramer,CGZ95,ZCG95,Zim96,ZC1,ZC2}. 
There is a kind of 
coherent length corresponds to the wave packet length 
scale of the emitter, which causes pions to concentrate at 
low momenta\cite{Zim96,ZC1,ZC2}.  Thus, it is very interesting to analyse the 
effects of multi-pion 
correlation and finite coherent length on two-pion interferometry\cite{Zim96,ZC1,ZC2}.

The general definition of the $n$ pion correlation function 
$C_{n}(\vec{p}_{1},\cdot \cdot \cdot \vec{p}_{n})$ is
\begin{eqnarray}
C_{n}(\vec{p}_{1},\cdot \cdot \cdot \vec{p}_{n})
=\frac{P_{n}(\vec {p}_{1},\cdot \cdot \cdot \vec{p}_{n})}
{\prod_{i=1}^n P_{1}(\vec{p_{i}})}   ,
\end{eqnarray}
where $P_{n}(\vec{p}_{1},\cdot \cdot \cdot \vec{p}_{n})$ is the 
probability of observing $n$ pions with momenta $\{ \vec{p}_{i} \}$
all in the same event.   The n-pion momentum probability 
distribution $P_{n}(p_1,\cdot\cdot\cdot p_n)$ 
 can be expressed as\cite{Zajc87,Pratt93,CGZ95,ZC1}
\begin{equation}
P_{n}(p_1,\cdot\cdot\cdot,p_n)
=\sum_{\sigma} \rho_{1,\sigma(1)}\rho_{2,\sigma(2)}...\rho_{n,\sigma(n)},
\end{equation}
with
\begin{equation}
\rho_{i,j}=\rho(p_{i},p_{j})=
\int d^{4}x g_w(x, \frac{(p_{i}+p_{j})}{2}) 
e^{i(p_{i}-p_{j})x}.
\end{equation}
$\sigma(i)$ denotes the $i$th element of a permutation of the sequence
${1,2,3,\cdot \cdot \cdot, n}$, and the sum over $\sigma$ denotes the 
sum over all $n!$ permutations of this sequence. $g_w(x,K)$ can be 
explained as the probability of finding a pion at point $x$ with momentum 
$p$ which is defined as\cite{Pratt84,CGZ94,SSS}
\begin{equation}
g_{w}(x,k)=\int d^{4}y j^{*}(x+y/2)j(x-y/2)exp(-iky) .
\end{equation}
Where $j(x)$ is the 
current of the pion, which can be expressed as\cite{CGZ94,ZCG95} 
\begin{equation}
j(x)=\int d^{4}x' d^{4}p j(x',p) \nu(x') exp(-ip(x-x')) .
\end{equation}
Here $j(x',p)$ is the probability amplitude of finding a pion 
with momentum $p$ , emitted by the emitter at $x'$. $\nu(x')$ 
is a random phase factor which has been taken away from $j(x',p)$.  
All emitters are uncorrelated in coordinate space when assuming:
\begin{equation}
<\nu^{*}(x')\nu(x)>=\delta^{4}(x'-x)   .
\end{equation}
This is in ideal case. In a more realistic case, each chaotic emitter 
has a small coherent wave packet length scale and 
the above 
equation can be replaced by\cite{Zim96}
\begin{equation}
<\nu^{*}(x')\nu(x)>=\frac{1}{\delta^{4}}
\exp\{-\frac{(x_{1}-x'_{1})^{2}}{\delta^{2}}
-\frac{(x_{2}-x'_{2})^{2}}{\delta^{2}}
-\frac{(x_{3}-x'_{3})^{2}}{\delta^{2}}
-\frac{(x_{0}-x'_{0})^{2}}{\delta^{2}} \}  .
\end{equation}
Here $\delta $ is a parameter which determines the coherent length (time) 
scale of the emitter.   For simplicity, the same coherent scale is taken 
for both spacial and time at the moment.  
The above formula shows that two-emitters within the range of 
$\delta$ can be seen as one emitter, while two-emitters out of this
range are incoherent.  For simplicity we 
also assume that
\begin{equation}
<\nu^{*}(x)>=<\nu(x)>=0    ,
\end{equation}
which means that for each emitter the phases are randomly distributed in the 
range of $0$ to $2 \pi$. Then we have the following relationship
\begin{equation}
<\nu^{*}(x')\nu^{*}(x)>=<\nu(x)\nu(x')>=0  .
\end{equation}

Inserting eq.(5) into eq.(4) we have
\begin{eqnarray}
g_{w}(Y,k)&=&\int d^{4}y exp(-iky)
\nonumber\\
	&&\int d^{4}x'j^{*}(x',p_{1})d^4p_{1}e^{ip_{1}(Y+y/2-x')}\nu^{*}(x')\\
	&&\int d^{4}x''j(x'',p_{2})d^4p_{2}e^{-ip_{2}(Y-y/2-x'')}\nu(x'') .
\nonumber
\end{eqnarray}
From eq.(1), the two-pion correlation function 
 can be expressed as
\begin{equation}
C_{2}(\vec{p}_{1},\vec{p}_{2})=
1+\frac{\int d^{4}x d^{4}x' g_{w}(x,k) g_{w}(x',k)
exp(iq(x-x'))}{\int d^{4}x d^{4}x'g_{w}(x,p_{1})g_{w}(x,p_{2})}  .
\end{equation}

For $n \pi$ events, the two-pion correlation function can be defined as
\begin{equation}
C_{2}^{n}(p_1,p_2)=\frac{P_{2}^{n}(\vec p_{1},\vec P_{2})}{P_{1}^{n}
(\vec p_{1})P_{1}^{n}(\vec p_{2})} ,
\end{equation}
where $P_{2}^{n}(\vec p_{1},\vec p_{2})$ and $P_{1}^{n}(\vec{p})$  is 
the modified 
two-pion and one-pion inclusive
distribution in $n$ pion events which can be expressed as
\begin{equation}
P_{2}^{n}(\vec p_{1},\vec p_{2})=\frac{\int \prod_{i=3}^{n} d\vec p_{i}
P_{n}(\vec p_{1}....\vec p_{n})}{\int \prod_{i=1}^{n} d\vec p_{i}
P_{n}(\vec p_{1}....\vec p_{n})},
\end{equation}
and
\begin{equation}
P_{1}^{n}(\vec p_{1})=\frac{\int \prod_{i=2}^{n} d\vec p_{i}
P_{n}(\vec p_{1}....\vec p_{n})}{\int \prod_{i=1}^{n} d\vec p_{i}
P_{n}(\vec p_{1}....\vec p_{n})}.
\end{equation}

Now we define the function\cite{Pratt93,CGZ95,ZC1}
\begin{equation}
G_{i}(p,q)= \int \rho(p,p_{1}) d\vec p_{1} \rho(p_{1},p_{2})
d \vec p_{2} \cdot \cdot \cdot \rho(p_{i-2},p_{i-1})d \vec p_{i-1}
\rho(p_{i-1},q).
\end{equation}

From the expression of $P_{n}(p_1,\cdot\cdot\cdot,p_n)$ 
\begin{equation}
P_{n}=\sum_{\sigma} \rho_{1,\sigma(1)}\rho_{2,\sigma(2)}...\rho_{n,\sigma(n)},
\end{equation}
the two-pion inclusive distribution can be expressed as
\begin{eqnarray}
P_{2}^{n}(\vec p,\vec q)=\frac{1}{n(n-1)}\frac{1}{\omega(n)}
\sum_{i=2}^{n}[\sum_{m=1}^{i-1}G_{m}(p,p)G_{i-m}(q,q)+G_{m}(p,q)
\cdot G_{i-m}(q,p)]\omega(n-i)
\end{eqnarray}
with
\begin{eqnarray}
\omega(n)=\frac{1}{n!}\int \prod_{k=1}^{n} d\vec p_{k} 
P_{n}(p_1,\cdot\cdot\cdot,p_n)~~~.
\end{eqnarray}
The single-pion distribution is
\begin{eqnarray}
P_{1}^{n}(\vec p)=
\frac{1}{n}\frac{1}{\omega(n)}\sum_{i=1}^{n}G_{i}(p,p)\omega(n-i).
\end{eqnarray}

From the expression of eq.(19), we have
\begin{eqnarray}
\omega(n)=\frac{1}{n}\sum_{i=1}^{n} C(i)\omega(n-i)
\end{eqnarray}
with
\begin{equation}
C(i)=\int d\vec p G_{i}(p,p).
\end{equation}

From the above method the two-pion and 
single pion inclusive distribution
can be calculated for $n$ pion events. 
In the follwoing, we will give an example to investigate the finite 
coherent length and multi-pion correlation effects on two-pion interferometry.  
We assume that the chaotic emitter amplitude distribution is
\begin{equation}
j(x,k)=\exp(\frac{-x_{1}^{2}-x_{2}^{2}-x_{3}^{2}}{2R_{0}^{2}})
\delta(x_{0})
exp(-\frac{k_{1}^{2}+k_{2}^{2}+k_{3}^{2}}{2\Delta_0^{2}})~~~.
\end{equation}
Where $R_0$ and $  \Delta_0 $ are the parameters which 
represents the radius of the chaotic source size and the 
momentum range of pions respectively.
Bringing  eq.(7) and eq.(22) into 
eq.(10), we can easily get the function $g_w(x,k)$ 
\begin{eqnarray}
g_w(x,k)= (\frac{1}{\pi R^2})^\frac{3}{2}
exp(-\frac{\vec{x}^2}{R^2})\delta(x_0)
(\frac{1}{\pi \Delta^2})^\frac{3}{2}
\exp\{-\frac{\vec{k}^{2}}{\Delta^2}\},
\end{eqnarray}
with 
\begin{eqnarray}
R^2=R_0^2+\frac{1}{\Delta_0^2},~~~ \frac{1}{\Delta^2}=\frac{1}{\Delta_0^2}+
\frac{R_0^2\delta^2}{\delta^2+4R_0^2} ~~~.
\end{eqnarray}

$g_w(x,\frac{p+q}{2})$ can be expressed as
\begin{equation}
g_w(x,\frac{p+q}{2})=
\frac{1}{(\pi R^{2})^{3/2}}e^{-\frac{r^{2}}{R^{2}}}  
\frac{1}{(\pi \Delta^{2})^{3/2}}e^{-\frac{(\vec{p}+\vec{q})^2}{4\Delta^{2}}}
\delta(t) .
\end{equation}
Then we have
\begin{equation}
\rho(p,q)=\int g_w(x, \frac{p+q}{2})e^{i(p-q)x}dx
=\frac{1}{(2\pi \Delta^{2})^{3/2}}
e^{-\frac{(p-q)^{2}R^{2}}{4}}e^{-\frac{(\vec{p}+\vec{q})^2}{4\Delta^{2}}}.
\end{equation}

Define
\begin{equation}
G_{n}(p,q)=\int \rho(p,p_{1}) 
\prod_{i=1,n-2} d\vec p_{i} 
\rho(p_{i},p_{i+1})
d\vec p_{n-1} 
\rho(p_{n-1},q)
\end{equation}
Using eq.(26),  we can easily get 
\begin{equation}
G_{n}(p,q)= \alpha_{n} e^{-a_{n}(p^{2}+q^{2})+g_{n} \vec p \cdot \vec q}
\end{equation}
where
\begin{equation}
a_{n+1}=a_1-\frac{g_1^2}{4b_{n}},~
b_{n}=a_{n}+a_1,~
g_{n+1}=\frac{g_{n}\cdot g_1}{2b_{n}}
\end{equation}
and
\begin{equation}
\alpha_{n+1}=\alpha_{n}(\frac{1}{\Delta^{2}})^{3/2}(\frac{1}{b_{n}})^{3/2}
\end{equation}
with 
\begin{equation}
a_{1}=\frac{R^{2}}{4}+\frac{1}{4\Delta^{2}}, ~ 
g_{1}=\frac{R^{2}}{2}-\frac{1}{2\Delta^2} , ~ 
\alpha_{1}=\frac{1}{(\pi \Delta^{2})^{3/2}}.
\end{equation}

The apparent  radius and coherent parameters derived from 
two-pion interferometry are defined as:
\begin{equation}
R_n=\sqrt{-\frac{\partial^2 C_2^n(q)}{\partial q^2}|_{q=0}},~~~
\lambda_n=C_2^n(q)|_{q=0}-1
\end{equation}
The apparent radius and coherent parameters derived from two-pion correlation 
are shown in Fig.1 and Fig.2. It is clear that 
 as the multiplicity of the event increases,  
the two-pion correlation function has a lower chaoticity, though
the actual source is totally chaotic and 
 the apparent 
radius derived from two-pion interferometry 
becomes smaller. 
From eq.(24), It can be seen 
clearly that as $\delta$ increases the effective temperature $\Delta$ becomes smaller, 
 the effect of multi-pion 
correlation on two-pion interferometry becomes larger. 
  From Fig.1 and Fig.2, we find that 
as the pion multiplicity increases, the BE correlation 
 and the coherent length  effects on two-pion interferometry becomes larger.  

Conclusions: In this paper, multi-pion Bose-Einstein correlation and 
the coherent length effects on two-pion interferometry are discussed.  
It was shown that multi-pion 
Bose-Einstein correlation and coherent length cause the apparent
radius and coherent parameter of source which derived from 
two-pion interferometry become smaller.  For larger pion multiplicity,
 the coherent length effects on the two-pion interferometry becomes 
 larger. It is also shown that as the coherent length becomes larger the 
 effective temperature becomes smaller.  

\begin{center}
{\bf Acknowledgement}
\end{center}

This work was partly supported by the Alexander von Humboldt foundation in 
Germany and National Natural Science Foundation of China.

\newpage
\begin{center}
{\bf Figure Captions}
\end{center}

\end {document}